\documentclass[a4paper,11pt]{article}
\pdfoutput=1 

\usepackage{jcappub} 
\usepackage{graphicx,subfigure}

\usepackage[T1]{fontenc} 

\title{Another look at redshift drift and the backreaction conjecture}

\author{S. M. Koksbang}
\affiliation{Department of Physics, University of Helsinki and Helsinki Institute of Physics, \\P.O. Box 64, FIN-00014 University of Helsinki, Finland}
\emailAdd{sofie.koksbang@helsinki.fi}


\abstract{
Earlier studies have conjectured that redshift drift is described by spatially averaged quantities and thus becomes positive if the average expansion of the Universe accelerates. This conclusion is reevaluated here by considering exact light propagation in a simple toy-model with average accelerated expansion. The toy-model and light propagation setup is explicitly designed for concordance between spatial averages and averages along light rays. While it is verified that redshift-distance relations are well described by average quantities in this setup, it is found that the redshift drift is not. Specifically, the redshift drift is negative despite the on-average late-time accelerated expansion of the model. This result implies that measuring redshift drift signals at low redshifts gives the potential for directly falsifying the backreaction conjecture. However, the results are based on a toy-model so it is in principle possible that the result is an artifact and that redshift drift is in reality well described by spatially averaged quantities. The result therefore highlights the importance of developing \emph{exact} solutions to the Einstein equations which exhibit average accelerated expansion without local expansion so that the relation between spatial averages and observations can be firmly established.
}
\begin{document}
\maketitle
\flushbottom

\section{Introduction}\label{sec:intro}
The next generation of observational data will be extremely precise and ample, permitting scrutinizing investigations into e.g. the detailed qualities of dark energy. The Large Synoptic Survey Telescope (LSST) is for instance expected to yield observational data from over 10 million supernovae \cite{lsst} while the Euclid mission is e.g. expected to measure the shapes of more than a billion galaxies \cite{euclid}.
\newline\indent
A particularly interesting feature of upcoming surveys is that they will provide the possibility of detecting effects hitherto too small to be measured. Specifically, real-time cosmology will become feasible. Real-time cosmology is the research field dedicated to studying the time variation of astrophysical observables in ``real time'' i.e. during observation time \cite{realtime}. Examples of real-time cosmology are cosmic parallax \cite{parallax1,parallax}, redshift drift \cite{Sandage,McVittie} and the recently proposed measurement of the drift of the polarization af inverse-Compton scattered CMB photons \cite{realtime_cmb}. Amongst the observables discussed in real-time cosmology, redshift drift has so far received the most attention and will also be the focus here. Redshift drift is a tiny quantity but the technological advances seen during the past few decades mean that measuring it is becoming practicable. The modern estimates of the observation time required to measure the redshift drift is typically of the order of a few decades \cite{liske} but some suggest that redshift drift could be measured before the year 2030 if dedicated instruments are built now \cite{5years}. Redshift drift measurements have therefore become a serious observational goal and is for instance the main application purpose of CODEX \cite{codex}, a spectrograph designed for the European Extremely Large Telescope (E-ELT). Redshift drift measurements with ELTs are expected to be in the redshift interval $2-5$ as the measurements are based on the Lyman $\alpha$ forest \cite{liske}. It is also expected that the redshift drift can be measured at redshifts below 1 e.g. by appropriately modifying the Square Kilometer Array (SKA) or similar instruments \cite{drift_with_SKA2,drift_with_SKA,ska_like,fast}.
\newline\newline
The redshift drift has the potential to become a highly valuable observable: Redshift drift measurements can significantly alleviate parameter degeneracies and be used to distinguish different cosmological models within standard cosmology \cite{realtime, forecast1, forecast2, forecast3, forecast4, forecast5, forecast6, forecast7, forecast8, forecast9} as well as distinguish between standard cosmological models and alternatives such as the timescape scenario \cite{timescape_observables} and varying speed of light cosmologies \cite{varying_c}. Redshift drift measurements can also be used to conduct principal investigations of fundamental assumptions underlying standard cosmology such as the Copernican principle \cite{Copernican, void,axial, void_mimic}. Lastly, and arguably most importantly, redshift drift measurements correspond to \emph{direct} and \emph{model independent} measurements of the cosmic expansion rate (or rather, its temporal variation). All in all, redshift drift is a highly important observable available to us in the near-future. It is therefore important to have a firm understanding of how different cosmological settings can affect redshift drift measurements. Specifically, since the Universe is not exactly homogeneous, redshift drift measurements will e.g. be prone to contamination for large-scale structures. For measurements based on the Lyman $\alpha$ forest, this contamination was in \cite{Loeb} estimated to be orders of magnitude smaller than the cosmic signal while it was in \cite{variance} through linear perturbation theory estimated to be as much as $\sim1\%$ for some redshifts with the possibility of reducing it to $\sim0.01\%$ by averaging over many signals. Complementary to this, \cite{peculiar_acc1, peculiar_acc2} study effects of peculiar acceleration of galaxies and galaxy clusters with the conclusion that the peculiar signal dominates over the cosmic signal at low redshifts (at least $z\lesssim0.5$) while it will be noise at higher redshifts, difficult to distinguish from the cosmic signal unless the cosmic signal can be obtained by averaging over many signals.
\newline\indent
Another important aspect of cosmological inhomogeneities is that they can affect the large-scale/average evolution of the Universe. This phenomenon is known as cosmic backreaction (see e.g. \cite{bc_review1,bc_review2,bc_review3,bc_review4} for reviews on the topic and section \ref{sec:toymodel} for a brief introduction). At this point, it is still up for debate to what extent cosmic backreaction is important in the real universe (see e.g. \cite{is_bc_important} and references therein). However, it is at least \emph{in principle} possible for backreaction to generate acceleration of the average expansion rate without there being any local accelerated expansion due to e.g. dark energy. The idea that backreaction is the source for the apparent late-time accelerated expansion of the Universe is known as the backreaction conjecture. Future observational data can indirectly give indications of the size of cosmic backreaction simply because it is unlikely that a universe with significant backreaction will behave very precisely as a $\Lambda$CDM model. Specifically, several FLRW (Friedmann-Lemaitre-Robertson-Walker) geometry consistency relations have been devised (e.g. \cite{sum_rule, parallax_test, copernican}) and the more precisely observations obey these, the less likely it is that backreaction is significant for the dynamics of the Universe. However, while a failure of observations to fulfill FLRW consistency relations directly falsifies standard cosmology, a precise observational fulfillment of FLRW consistency relations does not directly and unambiguously falsify the backreaction conjecture.
\newline\indent
It is desirable to devise observational tests that can directly falsify the backreaction conjecture. For this purpose, redshift drift might be a key. It has earlier been concluded that redshift drift is well described by spatially averaged quantities and hence that redshift drift will be positive at low redshifts regardless of whether the apparent late-time accelerated expansion is due to local or global acceleration \cite{dig_zdrift,av_obs2}. This conclusion is based on general, mainly statistical, considerations largely at a conjecture level and the conclusion is questioned here where the topic is studied anew by directly propagating light rays through a toy-model with average accelerated expansion. After a brief introduction to redshift drift in standard cosmology in section \ref{sec:FLRW}, the toy-model studied here is described in section \ref{sec:toymodel}. Different methods for describing average and exact redshift drift in the model are discussed in section \ref{sec:lightprop} and results obtained with these methods are given in both section \ref{sec:lightprop} and \ref{sec:results}. A summary is given in section \ref{sec:summary}.

\section{Redshift drift in FLRW models}\label{sec:FLRW}
The redshift drift $\delta z$ is defined as the change in the observed redshift of a (comoving) source during an observation time $\delta t_0$, i.e. $\delta z:=\frac{dz}{dt_0}|_{t_0}\delta t_0$. Since $\delta t_0$ is very small ($\sim 10-100$yrs) compared to cosmological time scales, the expression for the redshift drift can be simplified through a first order Taylor expansion. The expression is straightforward to derive from the FLRW expression for the redshift, $1+z = \frac{1}{a(t)}$, and can be written as
\begin{equation}\label{eq:dzFLRW}
\delta z = \delta t_0\left[ (1+z)H_0 - H(t_e)\right]  = \delta t_0 (1+z)\left[a_{,t}(t_0) - a_{,t}(t_e) \right] ,
\end{equation}
where $a(t_0) = 1$, subscripted commas indicate partial derivatives and a subscripted $0$ implies evaluation at observation time, and $e$ at emission time.
\newline\newline
The first expression for $\delta z$ given above is the typical rendering of the FLRW redshift drift. The second rendering of the expression highlights the fact that in an FLRW universe, the redshift drift is positive if the expansion accelerates between observation and emission time. This is for instance the case at late times in a standard $\Lambda$CDM model while it is specifically not the case in an FLRW solution to the Einstein equations containing only regular matter (radiation, baryonic matter, dust dark matter). For a coasting universe, the redshift drift vanishes identically which therefore yields a very strong prediction for e.g. the $R_h=ct$ universe \cite{melia}.

\section{Backreaction and toy-models with average accelerated expansion}\label{sec:toymodel}
One possible explanation of the apparent late-time accelerated expansion of the Universe is the backreaction conjecture which postulates that it is merely an artifact arising from interpreting observations with too simple cosmological models (the FLRW models). The backreaction conjecture is based on noting that the average evolution of inhomogeneous solutions to the Einstein equations generally deviate from FLRW evolutions. Since the Universe is not exactly spatially homogeneous and isotropic, this means that it may be inappropriate to use FLRW models as the base cosmological models when interpreting observations.
\newline\indent
The difference between FLRW evolution and the average evolution of a general inhomogeneous cosmological model is known as cosmic backreaction. Backreaction is typically studied by using Buchert's averaging scheme for scalars \cite{fluid1,bc_fluidII}. Within this scheme, averages on flow-orthogonal spatial hypersurfaces (vorticity is assumed zero) are defined as proper volume averages, i.e. $\left\langle S \right\rangle :=\frac{\int_{D}S\sqrt{|\det g_{ij}|}}{\int_{D}\sqrt{|\det g_{ij}|}}$, where $S$ is a scalar, $D$ is a spatial domain and $g_{ij}$ is the spatial metric. Applying Buchert's averaging scheme to the Hamiltonian constraint and the Raychaudhuri equation leads to the Buchert equations ($c=1$ throughout):
\begin{equation}\label{eq:Friedmann}
3H_D^2:=3\left( \frac{ a_{D,t}}{a_D}\right) ^2= 8\pi G\left\langle \rho\right\rangle - \frac{1}{2}\left\langle ^{(3)}R\right\rangle +\Lambda - \frac{1}{2}Q
\end{equation}
\begin{equation}\label{eq:acc}
3\frac{a_{D,tt}}{a_D} = -4\pi G\left\langle \rho\right\rangle + \Lambda + Q.
\end{equation}
The normalized volume scale factor $a_D$ is defined by $a_D:=\left( \frac{V_D}{V_{D_0}}\right)^{1/3} $ with $V_D$ the proper volume of the spatial averaging domain $D$ which should be chosen to be larger than the homogeneity scale. The source term $Q$ is the kinematical backreaction defined as $Q:=\frac{2}{3}\left(\left\langle \Theta ^2\right\rangle-\left\langle \Theta\right\rangle^2 \right) -2\left\langle \sigma^2\right\rangle $. $\Theta$ is the local fluid expansion rate and $\sigma^2:=\frac{1}{2}\sigma_{\mu\nu}\sigma^{\mu\nu}$ is the fluid's shear scalar.
\newline\newline
As noted in the introduction, the significance of cosmic backreaction for the real universe is a matter of great debate. The main problem is that it has proven very difficult to obtain a realistic quantification of backreaction because this requires a realistic, fully relativistic solution to the Einstein equations. Quantifications have e.g. been attempted with perturbation theory \cite{pert1,pert2,near-FLRW,light_cone_De1,light_cone_De2,light_cone_bc}, methods utilizing Newtonian N-body simulations \cite{multiscale,Nbody1,Nbody2,Nbody3}, relativistic simulations \cite{fitting1,relsim1,relsim2,relsim3,relsim4} and ensemble models \cite{ellipsoid, peak}. These methods all have shortcomings that make the resulting quantifications at most suggestive.
\newline\indent
Here, the goal is not to attempt quantifying backreaction. Instead, the focus is on the relation between spatially averaged quantities and observations in a universe with significant backreaction. In \cite{av_obs1,av_obs2} and \cite{template1,template2,template3} different methods for relating spatially averaged quantities to redshift and distance measures were proposed. These will be illustrated in the following with a toy-model, with the main interest being whether the redshift drift can be related to spatially averaged quantities as prescribed by these methods. This question has already been studied in \cite{dig_zdrift} where it was concluded that the redshift drift can indeed be described analogously to the FLRW case, using the volume scale factor and average Hubble rate instead of FLRW quantities. This conclusion is in line with the assertions in \cite{av_obs2} but the results were based on purely theoretical considerations without any numerical computations or examples to back them up - and the considerations did not involve an actual proof, only assertions. The former is remedied here, where the issue is studied by directly computing the redshift drift in a model with significant backreaction.

\subsection{Model setup}\label{sbsec:modelsetp}
For the purpose of the study, it is necessary to have a model of an inhomogeneous universe which exhibits average accelerated expansion at late times without having any local accelerated expansion. In addition, the model must be spatially statistically homogeneous and isotropic with a homogeneity scale of the order $\sim 100$Mpc. The latter features, expected to be true of the real universe, are necessary for light propagation studies; if structures become excessively large, a light ray will not experience a statistically homogeneous and isotropic universe because structures evolve too much during the time it takes a light ray to traverse the (spatial) homogeneity scale. Unfortunately, exact solutions to the Einstein equations with these features do not currently exist so a toy-model must be constructed instead. Specifically, a cosmological model with the desired features can be constructed by considering an ensemble of disjoint FLRW regions. Ensembles of disjoint regions have earlier been studied in relation to the backreaction problem, e.g. in \cite{Bull_tworegion, boehm, peak, ellipsoid, Syksy_2region, Syksy_2region2, Syksy_2region3} with different levels of complexity. For the current work, it does not matter how many different FLRW regions are included in the ensemble. Therefore, instead of considering a complicated ensemble of structures of many scales as in e.g. \cite{peak}, simply using two different FLRW models will suffice. The model considered here thus consists of an ensemble of two types of FLRW regions with a spatial arrangement mimicking an inhomogeneous universe containing a distribution of one type of voids and one type of overdensities. A similar interpretation of this type of model was given in e.g. \cite{boehm} and the model can be considered a simple version of a multiscale model \cite{multiscale,multiscale2}. Since the regions are disjoint, spatial averaging can be conducted by considering only one of each region, but using the appropriate volume fraction of the two regions. Thus, when considering spatial averages of the model, it is appropriate to simply use the formalism for two-region models considered in e.g. \cite{Syksy_2region,Syksy_2region2,Syksy_2region3}. This type of two-region models have earlier been used for studying the relation between backreaction and observations (e.g. \cite{Bull_tworegion,use2region,dig_zdrift}) and form part of the basis for computations in the timescape scenario (see e.g. \cite{timescape_observables} for a comparison of timescape predictions with observations).
\newline\newline
The volume scale factor of a two-region model and its corresponding average expansion rate are given by
\begin{equation}
a_{D}  = \left( \frac{a_u^3 + a_o^3}{a_{u,0}^3 +a_{o,0}^3 }\right) ^{1/3}
\end{equation}
\begin{equation}
H_{D}  = H_u\frac{a_u^3}{a_u^3+a_o^3} + H_o\frac{a_o^3}{a_u^3+a_o^3}.
\end{equation}
Subscripts $o$ and $u$ identify the Hubble parameters and scale factors of the two different FLRW models. These specific subscripts are chosen to indicate that one FLRW model is underdense while the other is overdense compared to an Einstein-de Sitter (EdS) universe.
\newline\indent
The two-region model studied here is based on an empty underdense region and an overdense region described by a closed, dust-only FLRW model. In this case, the development angle $\phi$ of the overdense region can be used to describe the relation between the scale factors and cosmic time as follows:
\begin{equation}
\begin{split}
t &= t_0\frac{\phi-\sin(\phi)}{\phi_0-\sin(\phi_0)}\\
a_u &= \frac{f_u^{1/3}}{\pi}\left( \phi - \sin(\phi)\right)\\
a_o &= \frac{f_o^{1/3}}{2} \left( 1-\cos(\phi)\right).
\end{split}
\end{equation}
$f_o$ and $f_u = 1-f_o$ are the volume fractions of the over- and underdense regions at $\phi = \pi$. The model studied here is defined as in \cite{dig_zdrift} so that at present time $\phi = \phi_0 = 3/2\pi$ and $H_{D_0}=70$km/s/Mpc. The parameter $f_o$ can be varied in order to obtain different average expansion histories. Further details regarding the computation of quantities in the two-region models are given in appendix \ref{app:2region}.
\newline\indent
Note that setting $H_{D_0}=70$km/s/Mpc is simply done for the obvious aesthetic reason of having an average Hubble constant close to the observed value. The actual value chosen for $H_{D_0}$ has no significant consequences for the results. Similarly, choosing present time to be at $\phi_0 = 3/2\pi$ is simply done because it corresponds to a time somewhat after the overdense region has begun contracting, i.e. it corresponds to having collapsing structures at present time (without the structures having collapsed too much compared to a more realistic scenario where complete collapse is avoided through virialization). Again, the specific choice of $\phi_0$ has no actual consequences for the conclusions of the work. The same is true for choosing the underdense region to be completely empty. It could just as well have been chosen to be a dust+curvature FLRW model with large negative curvature. Using an empty region is done for simplicity.
\newline\newline
\begin{figure*}
\centering
\subfigure[]{
\includegraphics[scale = 0.7]{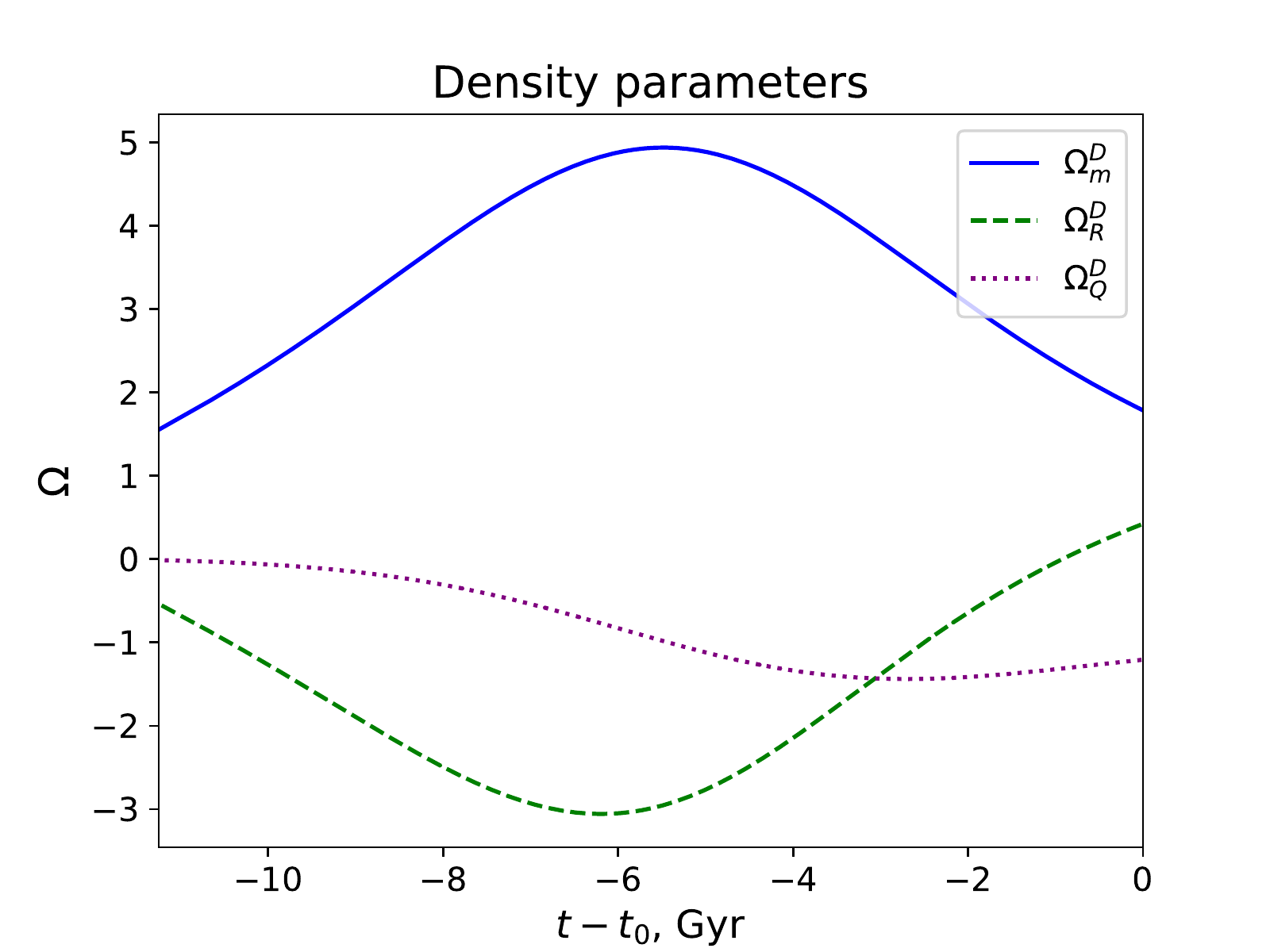}
}
\caption{Average density parameters of the studied two-region model.}
\label{fig:omegas}
\end{figure*}
Results from one particular model will be studied, but the results are in agreement with those found when using other particular models, obtained by varying $f_o$. The results are shown for a model defined by having $f_o = 0.5$. This choice is made somewhat arbitrarily but with the following considerations in mind: In order to obtain accelerated expansion of the size similar to what is actually observed, only $f_o\approx 0.18$ is necessary, but a larger value of $f_o$ will increase the late-time average accelerated expansion and will thus highlight any effects of this.
\newline\newline
The density parameters of the model are shown in figure \ref{fig:omegas} and are defined as follows:
\begin{equation}\label{eq:Omegas}
\begin{split}
\Omega^D_{m} :&=\frac{8\pi G\left\langle \rho\right\rangle _D}{3H_D^2}\\
\Omega^D_R:&=-\frac{\left\langle ^{(3)}R\right\rangle _D}{6H_D^2}\\
\Omega^D_\Lambda:&=\frac{\Lambda}{3H_D^2}\\
\Omega^D_Q:&=-\frac{Q_D}{6H_D^2}
\end{split}
\end{equation}
With these definitions, equation \eqref{eq:Friedmann} implies that these quantities add to 1, i.e.
\begin{equation}
\Omega^D_m + \Omega^D_R + \Omega^D_\Lambda+\Omega^D_Q = 1.
\end{equation}
Note that $\Omega^D_R$ and $\Omega^D_Q$ are negative when $\left\langle ^{(3)}R\right\rangle _D$ and $Q$ are positive and that $\Omega^D_{\Lambda} = 0$ here. It is clearly seen in the figure that there is significant kinematical backreaction, with $\Omega_Q^D$ of similar size as $\Omega_m^D$ and $\Omega^D_R$ at late times. The kinematical backreaction is indeed so significant that it induces a sign change of the curvature during the model's evolution (a sign change of the spatial curvature is not possible in a pure FLRW model and is thus necessarily a backreaction effect).
\begin{figure*}
\centering
\subfigure[]{
\includegraphics[scale = 0.7]{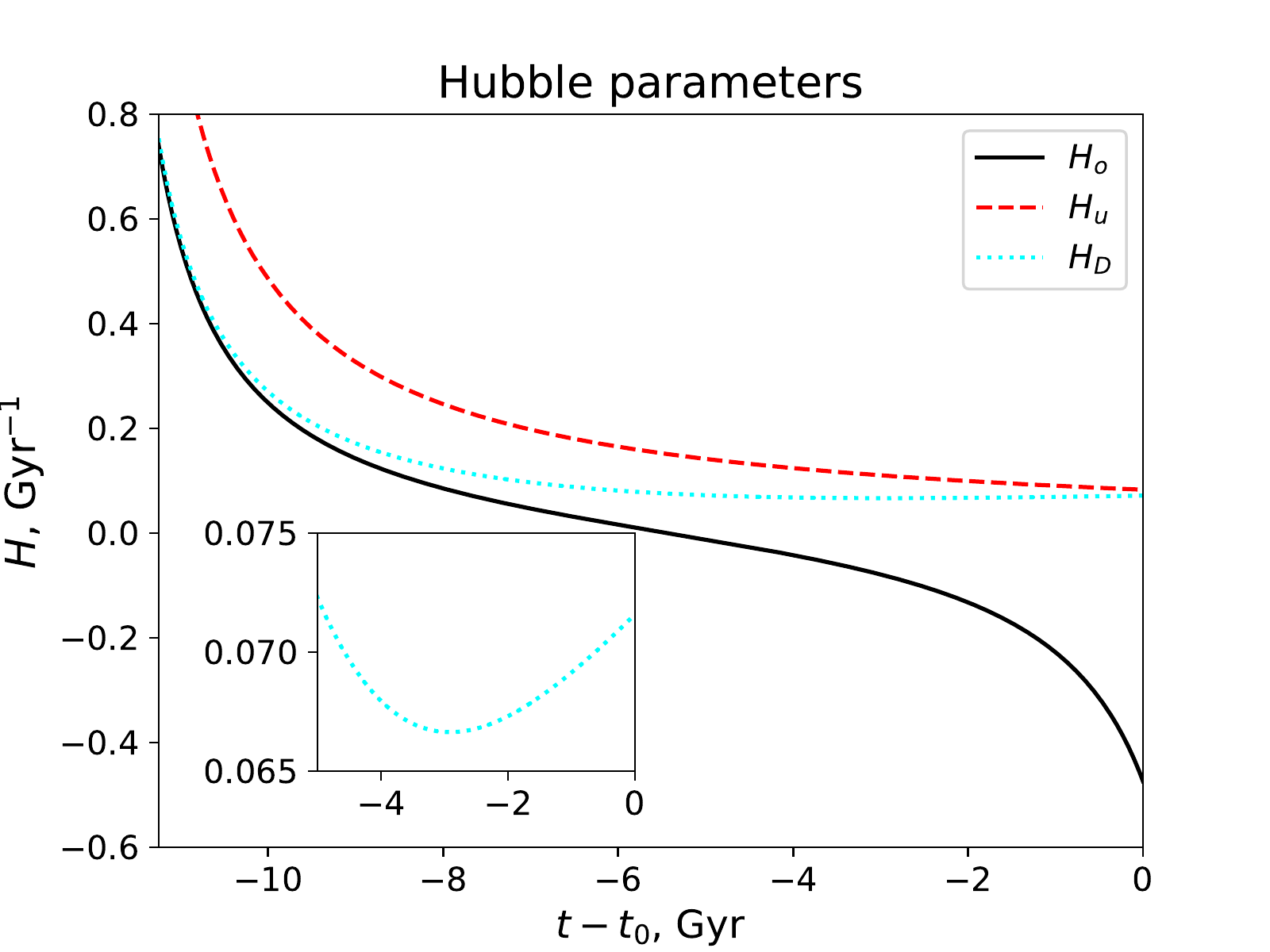}
}
\caption{Hubble parameters of the studied two-region model. A close-up of the average Hubble parameter, $H_D$, is included in order to show that it increases at late times.}
\label{fig:H}
\end{figure*}
\newline\newline
The three Hubble parameters $H_o, H_u$ and $H_{D}$ are shown in figure \ref{fig:H}. At early times, when the overdense regions make up the larger volume fraction, $H_D\approx H_o$ while at later times $H_D\approx H_u$. $H_o$ and $H_u$ are strictly decreasing functions, with $H_o<H_u$ at present time. The average Hubble rate, however, increases at late times, implying average accelerated expansion.

\section{Light propagation in the toy-model}\label{sec:lightprop}
This section describes exact schemes for computing the redshift drift in the toy-model as well as schemes proposed in the literature for relating average quantities to observations in a statistical/on-average way.

\subsection{Exact light propagation}\label{subsec:exactlight}
Exact light propagation in the toy-models is studied by solving the ordinary first order differential equations corresponding to the geodesic equation and the transport equation in the FLRW limit (for details on the latter, see e.g. \cite{arbitrary_spacetime}). With $s$ denoting an affine parameter and $D$ the distortion matrix, the latter is given by
\begin{equation}\label{D_dot}
\frac{d^2}{ds^2} D^a_b = T^a_cD^c_b,
\end{equation}
where $T$ is the optical tidal matrix given by
\begin{equation}
 T_{ab} = 
  \begin{pmatrix} \mathbf{R}- Re(\mathbf{F}) & Im(\mathbf{F}) \\ Im(\mathbf{F}) & \mathbf{R}+ Re(\mathbf{F})  \end{pmatrix}.
\end{equation}
$\mathbf{R}: = -\frac{1}{2}\mathcal{R}_{\mu\nu}k^{\mu}k^{\nu}$ and $\mathbf{F}:=-\frac{1}{2}C_{\alpha\beta\mu\nu}(\epsilon^*)^{\alpha}k^{\beta}(\epsilon^*)^{\mu}k^{\nu}$, with $\mathcal{R}_{\mu\nu}$ the Rici tensor, $C_{\alpha\beta\mu\nu}$ the Weyl tensor and $\epsilon^{\mu} := E_1^{\mu} + iE_2^{\mu}$ with $E_1^{\mu}, E_2^{\mu}$ spanning screen space. The Weyl tensor vanishes in FLRW spacetimes and the Einstein equations combined with the null-condition of the tangent vector $k^{\mu}$ yields $\mathbf{R} = -4\pi G\rho \left( k^t\right) ^2$. The transport equation is therefore quite simple in the FLRW limit. Further simplifications are obtained by noting that the light rays can be propagated with only $k^t$ and $k^r$ non-vanishing.
\newline\newline
The crucial consideration regarding light propagation in the toy-model is that light rays must be propagated alternately through the two types of FLRW regions with a ratio between the two corresponding to the proper volume fractions of the total ensemble (i.e. of the corresponding two-region model). The rate of alteration must correspond to a homogeneity scale of reasonable size. The requirement that the ratio between the proper lengths of the two types of regions along a light ray must equal the ratio of their proper volumes in the total ensemble implies that $l_o = l_u\left( \frac{a_o}{a_u}\right) ^2$, where $l_u$ and $l_o$ are, respectively, the comoving lengths of the under- and overdense regions traversed by a light ray before it is moved to the other type of region. The overall homogeneity scale is fixed by setting $l_u = 50$Mpc which corresponds to a present time size of underdense regions of approximately $72$Mpc and a present time size of overdense regions of approximately $20$Mpc. Increasing $l_u$ and $l_o$ means that neighboring disjoint regions will not ``compensate'' each other as well along a light ray because of evolution of the regions during the light ray travel time. This leads to an accumulating offset between average and exact results.
\newline\indent
Note that the choice $l_o = l_u\left( \frac{a_o}{a_u}\right) ^2$ is based on the volume fractions of the two types of regions and not a line-of-sight average. In Euclidean space this is a justifiable choice; if space is statistically isotropic 3D and 1D spatial averages should be identical and if structures evolve reasonably slowly, the average along a light ray can be approximated as a 1D spatial average. As noted in \cite{Tardis}, the relation between 1D and 3D spatial averages in curved spacetime is less trivial but this will not be considered further here even though the model studied here is neither locally nor globally/on-average spatially flat. With the choice of $\frac{l_o}{l_u}$ used here, average quantities along light rays will follow spatial averages by design.
\newline\newline
Important points will be illustrated with the redshift-distance relation, but the main point is to study the redshift drift which can be computed by solving the set of equations
\begin{equation}
\begin{split}
\frac{dt}{dr} &= -a\\
\frac{dz}{dr} &= (1+z)a_{,t}\\
\frac{d\delta z}{dr} &= a_{,t}\delta z + (1+z)a_{,tt}\delta t\\
\frac{d\delta t}{dr} &= -a_{,t}\delta t.
\end{split}
\end{equation}
These equations are straightforward to derive for a radial light ray in an FLRW model using a procedure similar to e.g. the derivations in section 3 of \cite{marie-noelle_sn}, i.e. by considering a light ray observed at $t = t_0$ and one at $t = t_0+\delta t_0$ and Taylor expanding to obtain a simple expression for the difference. One may note that a (slightly less sophisticated) method for computing the redshift drift along repeatable light paths is to simply compute the redshift along a light ray for the same observer at $t = t_0$ and $t = t_0+\delta t_0$ and subtract the two results. A light ray's path is repeatable if a light ray emitted from the same point of emission at different times follow the same spatial path \cite{rlp}. All light rays in FLRW spacetimes are repeatable.

\subsection{Average light propagation}\label{subsec:avlight}
Cosmic backreaction is typically described using spatial averages. However, as pointed out many times before (e.g. \cite{av_obs1,av_obs2,template1,fitting1,fitting2,fitting3,lightcone_dz,light_cone_bc}), a possible issue with such a procedure is that observations are mainly made on the light cone. Therefore, spatially averaged quantities (including backreaction terms) are only relevant if they can be related to observables in a sensible manner. Two specific schemes for relating spatial averages with observations have been proposed and used in the literature. One scheme can be characterized as a ``covariant'' scheme and was proposed in \cite{av_obs1,av_obs2} (but see e.g. also \cite{linder} for early considerations). The other scheme is based on introducing a so-called template metric and was proposed in \cite{template1} (but see e.g. also \cite{template2,template3}). The two schemes are described and compared in \cite{dig_selv_tardis} and here only a brief summary of the methods will be given.
\newline\newline
The covariant scheme utilizes considerations of statistical homogeneity and isotropy to conclude that the redshift and angular diameter distance, $D_A$, upon averaging over many data points should be well described by the pair $\left(z^C, D_A^C \right) $ given by
\begin{equation}
\begin{split}
1+z^C& = \frac{1}{a_D} \\
H_D\frac{d}{dz^C}\left(\left(1+z^C \right) ^2H_D \frac{dD_A^C}{dz^C}\right)& = -4\pi G\left\langle \rho\right\rangle_D D_A^C.
\end{split}
\end{equation}
The analysis leading to the covariant scheme relies on the requirement that spatial averages are computed on hypersurfaces of statistical homogeneity and isotropy and assumes that structure evolution is slow compared to the time it takes a light ray to traverse the homogeneity scale. These assumptions must therefore also be required by the toy-model as already noted.
\newline\newline
The template scheme is based on Ricci flow normalization and the introduction of a ``template'' metric $ds^2_D=-dt^2 + a_D^2\left(\frac{dr^2}{1-k_D(t)^2} + d\Omega^2 \right)$ based on an on-average description of null-geodesics. The resulting pair of observables $\left(z^{T1},D_A^{T1} \right) $ is given by
\begin{equation}\label{eq:temp1}
\begin{split}
\frac{dr}{dt}&=\frac{1}{a_D}\sqrt{1-k_D(t)r^2}\\
\frac{dk^t}{dt}&=\frac{1}{2}k^t\left(2H_D - \frac{k_{D,t}r^2}{1-k_Dr^2} \right)\\
D_A^{T1} &= r(z^T)a_D(z^{T1})\\
1+z^{T1}&=\frac{\left( k_{\mu}u^{\mu}\right) _e}{\left(k^{\nu}u_{\nu} \right) _0} = \frac{\left( k^t \right) _e}{\left( k^t \right) _0},
\end{split}
\end{equation}
where the choice $k_D(t)=\frac{1}{6}a_D^2\left\langle ^{(3)}R\right\rangle $ is assumed.
\newline\newline
As noted in \cite{dig_selv_tardis}, there is some ambiguity within the template scheme as there does not seem to be an {\em a priori} reason to prefer the template metric given in the above coordinates over one e.g. given according to $ds^2=-dt^2+a_D^2\left(dw^2+\frac{1}{k_D}\sin^2\left(w\sqrt{k_D} \right)d\Omega^2\right) $. This version of a template metric was e.g. used in  \cite{virialisation}. According to this choice, the redshift-distance pair, $\left(z^{T2}, D_A^{T2} \right) $, can be computed according to
\begin{equation}\label{eq:temp2}
\begin{split}
\frac{dw}{dt}& = -\frac{1}{a_D}\\
\frac{dk^t}{dt} &= H_Dk^t\\
D_A^{T2} &= \frac{a_D}{\sqrt{k_D}}\sin\left(w \sqrt{k_D}\right)\\
1+z^{T2}&= \frac{\left( k^t \right) _e}{\left( k^t \right) _0}.
\end{split}
\end{equation}
When $k_{D,t}\neq 0$, this set of equations yields different results than the set in Eq. \eqref{eq:temp1}. This is particularly clear for the expressions for $k^t$ and hence the redshift.

\subsection{Redshift-distance relations in the toy-model}
\begin{figure*}
\centering
\includegraphics[scale = 1]{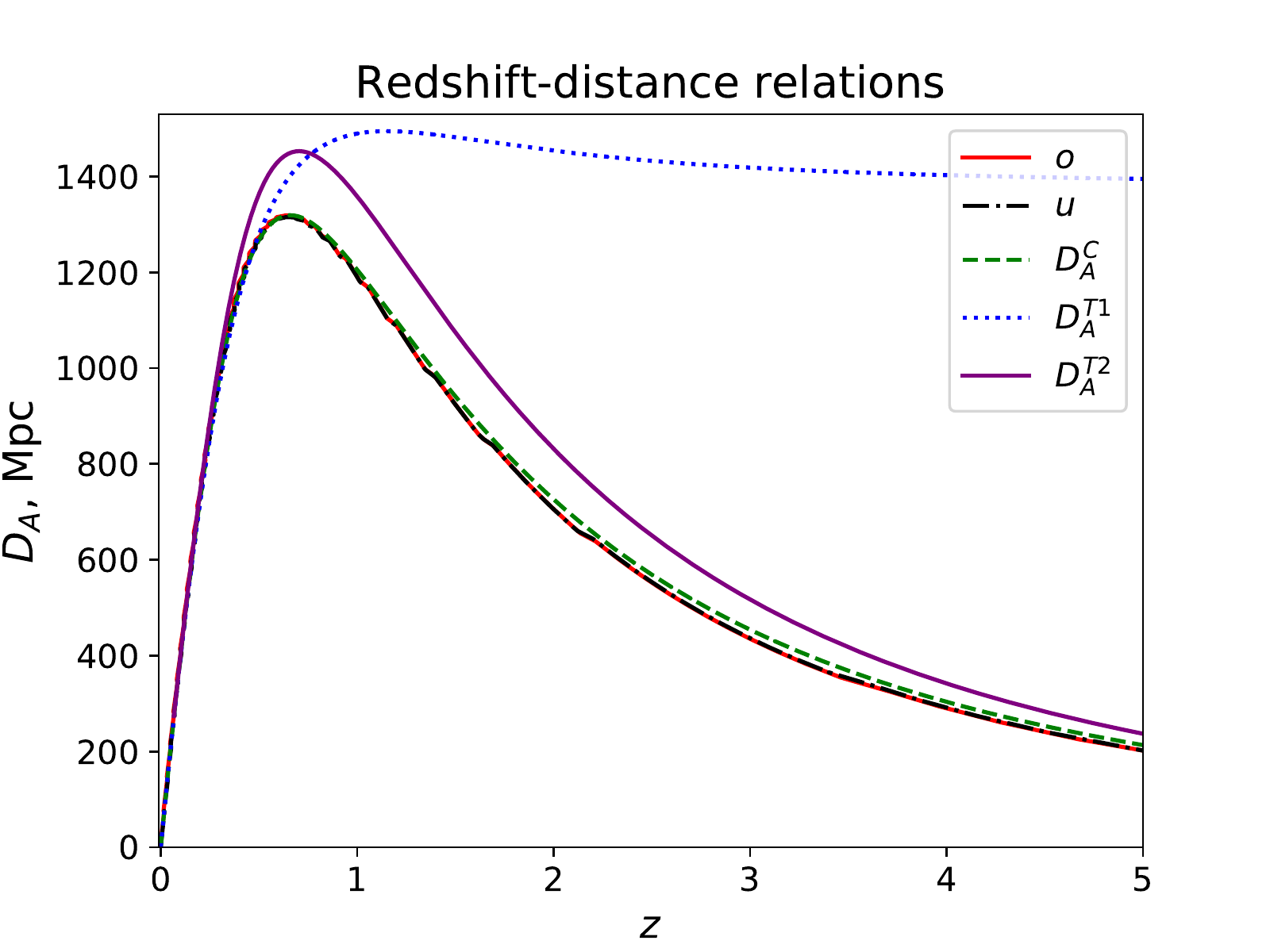}
\caption{Redshift-distance relations in toy-model. Results obtained by propagating light through disjoint regions are labeled according to whether the observer was placed in an empty FLRW region (``u'') or the overdense FLRW model (``o''). In both cases, the light ray was propagated the length $l_u$ ($l_o$) before being moved into a new region. }
\label{fig:DA}
\end{figure*}
Determining which, if any, of the above relations give good descriptions of observations upon averaging over many data points requires studying light propagation in statistically homogeneous and isotropic exact solutions to the Einstein equations with significant cosmic backreaction without large local effects. No such models are currently known. However, there is a preferred redshift-distance relation to use for the toy-model considered here as the model is specifically designed to fulfill the postulates leading to the pair $\left(z^C, D_A^C \right) $. In figure \ref{fig:DA}, the exact redshift distance relation for the toy-model is shown together with the three ``average'' descriptions given above. It is clearly seen that the covariant scheme gives an excellent description of the exact redshift-distance relation. This is in agreement with the findings of \cite{Bull_tworegion} where models similar to the one introduced here were studied in terms of different measures of acceleration. On the other hand, figure \ref{fig:DA} also shows that the two template schemes do not give good approximations of the exact redshift-distance relation - especially not the pair $\left(z^{T1},D_A^{T2} \right) $. The difference between the predictions of the two pairs $\left(z^{T1}, D_A^{T1} \right) $ and $\left( z^{T2}, D_A^{T2}\right) $ is also noteworthy as it emphasizes the possible issue with the template scheme in its current form.
\newline\indent
As discussed in the previous subsection, an accumulating off-set between exact and average results can be expected if structures and hence the homogeneity scale is large. This is seen in figure \ref{fig:DA} between the exact results and $\left( D_A^C,z^C\right) $. The exact results are shown both with the observer placed in an underdense and overdense region but the results depend only very little on this choice. The off-set is substantially decreased by scaling $l_u$ and $l_o$ by a factor of $0.1$ and somewhat increased by scaling them by a factor of $10$.
\newline\newline
Since it is only the covariant scheme that yields a good description of the average redshift-distance relation, it is only this scheme that will be used to study average redshift drift. Note however that $z^{2T} = z^C$. Note also that since the toy-model studied here is not an exact solution and is constructed specifically to fulfill the criteria leading to $\left( z^C, D_A^C\right) $, the results shown in figure \ref{fig:DA} cannot be considered a ``falsification'' of the template scheme. At the very least, however, figure \ref{fig:DA} shows that for this particular type of toy-model, the covariant scheme is the more appropriate one to use.
\newline\indent
With $z^C = \frac{1}{a_D}-1$, the average redshift drift is given by an expression analogous to the FLRW expression, i.e.
\begin{equation}
\delta z^C = \left[ (1+z)H_{D_0} - H_D(t_e)\right] \delta t_0.
\end{equation}
In the following subsection, this expression is compared to the exact redshift drift of the toy-model.

\section{Redshift drift in the toy-model}\label{sec:results}
\begin{figure*}
\centering
\includegraphics[scale = 1]{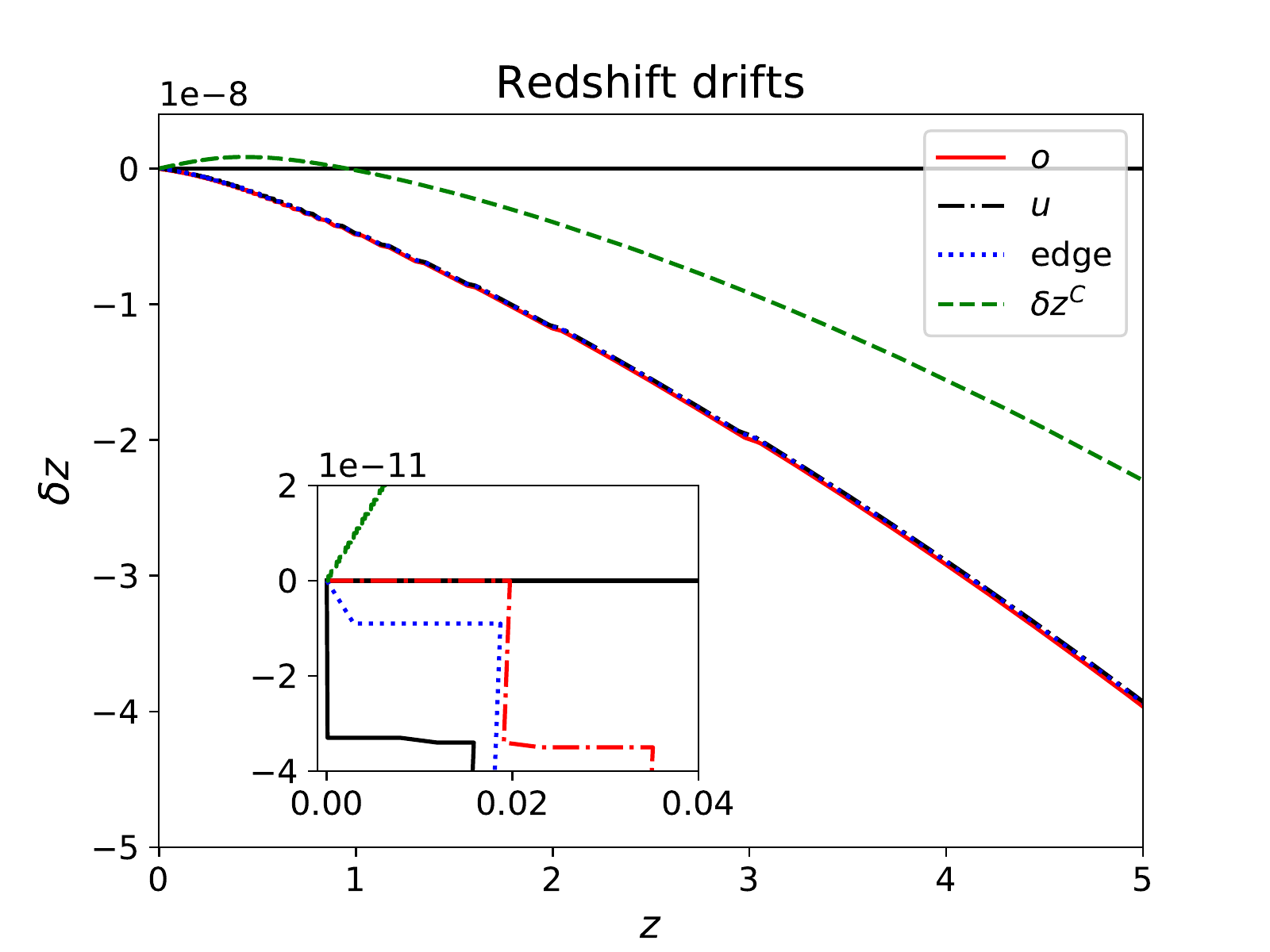}
\caption{Redshift drift in toy-model. Results obtained by propagating light through disjoint regions are labeled according to whether the observer was placed in an empty FLRW region (``u'') or an overdense FLRW model (``o''). In both cases, the light ray was propagated the length $l_u$ ($l_o$) before being moved into a new region. The light ray labeled ``edge'' corresponds to an observer placed approximately $1$Mpc from the edge of an overdense region so that the light ray was moved into an underdense region quickly after being initialized (at the point of observation). A close-up of the region $z\approx 0$ is included to highlight that the redshift drift is non-positive at low redshifts for the exact computations. The results were obtained with $\delta t_0 = 30$yrs. }
\label{fig:dz}
\end{figure*}

The redshift drift results obtained from propagating light rays in the toy-model are shown in figure \ref{fig:dz}. The exact redshift drift is shown for three different observers: One placed in an overdense region, one placed in an underdense region and one placed at the edge of an overdense region propagating into an underdense region after only roughly 1 Mpc. As seen, the main features of the exact redshift drift do not depend on the position of the observer.
\newline\newline
It is clear from the figure that the covariant scheme does not capture the overall behavior of the exact redshift drift. Most notably, the exact redshift drift is non-positive for all redshifts. It is a very important result that the redshift drift does not appear to become positive simply due to average accelerated expansion as it gives a possibility to attempt a direct falsification of the backreaction conjecture by measuring the redshift drift at low redshifts where the $\Lambda$CDM model predicts a positive redshift drift. One caveat is that peculiar redshift drifts might be positive and as mentioned in the introduction, these are expected to dominate over the cosmic redshift drift at least until $z\gtrsim0.5$ (but depending on the redshift source, this might not have direct impact on measurements of redshift drift). Positive redshift drifts due to local effects other than local accelerated expansion have also been illustrated e.g. in \cite{void} with a ``giant hump'' LTB model.
\newline\newline
It must be emphasized that the model studied here is not an exact solution to the Einstein equations and the result of a non-positive redshift drift must therefore be considered with some caution; it may be that the result is an artifact of the individual regions of the toy-model being disjoint. Indeed, it must be expected that much information is lost when removing smooth transitions between over- and underdense regions. For instance, in \cite{teppo1,teppo2}, comparisons of LTB models and two-region models showed that the disjoint two-region models exhibited significantly larger kinematical backreaction than the LTB models where the kinematical backreaction was suppressed by shear. Similarly, light propagation could in principle be significantly altered if the regions of the disjoint FLRW models studied here were joined in a smooth manner (irrespective of the change this would give in spatial averages). Note for instance that it was found in \cite{Tardis,dig_CMB} that the contributions of the shear and of the fluctuations in the expansion rate cancel with each other but not individually along light rays in specific LTB and Szekeres models. A similar result was found through analytical considerations in \cite{near-FLRW} for near-FLRW spacetimes. However, as recently shown in \cite{dig_selv_tardis}, the cancellation of the projected shear and expansion rate fluctuations is not a generic feature of light rays in LTB models. In the model studied here however, the fluctuations in the expansion rate cancel along the light ray. Disregarding the cosmic evolution on the time scales of a light ray's traversal of the homogeneity scale of the model, the cancellation is exact by design.
\newline\indent
Since the result presented here was not obtained with an exact solution to the Einstein equations, it cannot be excluded that $\delta z^C$ will turn out to give a good description of the average redshift drift in an exact solution with similar average features. If this is the case, $\delta z^C$ would be the more appropriate quantity to use for studying redshift drift in backreaction scenarios, including toy-models such as the one studied here and in \cite{dig_zdrift}, and positive redshift drift will appear due to average accelerated expansion, not only local accelerated expansion.
\newline\newline
With the above in mind, it is still striking that the average redshift and redshift-distance relation are well reproduced (up to small fluctuations) by the covariant approach while only the redshift drift is not. The reason that the redshift drift is not well described by the spatially averaged quantities may be related to the fact that the redshift drift is the derivative of an integral expression. Indeed, in order to obtain the expression for $z^C$ it is crucial that the redshift can be expressed as an integral along the light ray. Once the differential is included, it becomes less trivial to predict the average behavior of the quantity. This is especially true for quantities as small as the redshift drift where even small temporal variations in $z^C-z$ can have a big impact. This possible explanation suggests that real-time observables in general might be poorly described by spatially averaged quantities if there is significant backreaction. If this is indeed the case, it will be necessary to introduce more sophisticated approaches for describing this type of observations. One possible approach is that based on light cone averages introduced in \cite{lightcone_dz} where averaging is done on light cones constructed by using geodesic light cone coordinates (reminiscent of observational coordinates \cite{obs_coord} but not identical to these as explained in e.g. the introduction of \cite{lc_not_obs}). The procedure has e.g. been used in combination with perturbation theory \cite{light_cone_De1,light_cone_De2,light_cone_bc} and in \cite{lightcone_dz} it was used to derive an expression for the redshift drift averaged on a geodesic observer's light cone for a generic inhomogeneous universe. In a sense, such a setup is the opposite of that studied here: Here, the purpose is understanding aspects of the relation between spatial averages and observations while with schemes based on light cone averages, a main challenge is to understand how to relate light cone averages to cosmological models which are typically described and understood in terms of foliations with hypersurfaces of fixed time coordinate.

\section{Summary}\label{sec:summary}
A toy-model of disjoint FLRW regions (positively curved and void) was constructed in a way that resulted in late-time average accelerated expansion without any local accelerated expansion due to e.g. dark energy. Light propagation in the model was studied both by exact ray tracing through an ensemble of disjoint regions and through schemes proposed in the literature for describing on-average light propagation in statistically homogeneous and isotropic spacetimes through spatially averaged quantities. The model was specifically designed to fulfill the assumptions leading to one of these schemes (the covariant scheme). The covariant scheme was the only one that gave a good description of the exact redshift-distance relation. Therefore, only this scheme was used to study redshift drift in the model. According to this scheme, the redshift drift should be positive at low redshifts in the studied model due to the late-time average accelerated expansion. However, the exact redshift drift computed along light rays with different observers was non-positive at all redshifts. This result must be considered with caution as it was obtained by studying a toy-model that was specifically not an exact solution to the Einstein equations. It is striking though, that the studied model was specifically designed for the assumptions of the covariant scheme to be fulfilled and that the redshift and redshift-distance relation were both well reproduced by the covariant scheme, while only the redshift drift was not.
\newline\indent
If the result obtained here is valid in general for models exhibiting non-negligible backreaction, it gives a possible method to attempt a falsification of the backreaction conjecture which proposes that the apparent late-time accelerated expansion is an effect of averaging and not an actual, local acceleration. It is therefore important that exact solutions to the Einstein equations exhibiting average but not local accelerated expansion are developed so that it can be firmly established whether or not redshift drift becomes positive due to average accelerated expansion.

\acknowledgments
The author thanks Syksy Rasanen for comments on the presented results as well as the anonymous referee whose comments have led to significant improvements of the manuscript. The author is supported by the Independent Research Fund Denmark under grant number 7027-00019B.

\appendix
\section{Two-region models}\label{app:2region}
This appendix details computations of quantities of the two-region models studied in the main text. Some of the information is also given in e.g. \cite{Syksy_2region,Syksy_2region2,Syksy_2region3}.
\newline\newline
In a positively curved dust-only universe, the time coordinate $t$ is related to the development angle simply by $t\propto \phi - \sin(\phi)$ while the scale factor is $a_o\propto \left( 1-\cos(\phi)\right) $. The scale factor in an empty FLRW model is $a_u\propto t\propto \phi - \sin(\phi)$ \cite{Ryden}. Thus, the relation between time and the two scale factors of the studied two-region model is (as given in the main text):
\begin{equation}
\begin{split}
t = t_0\frac{\phi-\sin(\phi)}{\phi_0-\sin(\phi_0)}\\
a_u = \frac{f_u^{1/3}}{\pi}\left( \phi - \sin(\phi)\right)\\
a_o = \frac{f_o^{1/3}}{2} \left( 1-\cos(\phi)\right)
\end{split}
\end{equation}
Following \cite{Syksy_2region}, the average Hubble and deceleration parameters are given by (a subscripted $2region$ will be used her instead of $D$ to emphasize that the results are valid only for averages over a two-region model):
\begin{equation}
H_{2region} = H_o\left( 1-v+vh\right) 
\end{equation}
\begin{equation}
\begin{split}
q_{2region}  &= q_u\frac{1-v}{\left(1-v+hv \right) ^2} + q_o\frac{vh^2}{\left(1-v+hv \right) ^2}-2\frac{v\left( 1-v\right) \left(1-h \right)^2 }{\left(1-v+hv \right) ^2} \\
&= q_o\frac{vh^2}{\left(1-v+hv \right) ^2}-2\frac{v\left( 1-v\right) \left(1-h \right)^2 }{\left(1-v+hv \right) ^2}
\end{split}
\end{equation}
The auxiliary functions $h$ and $v$ used above are defined by $h:=\frac{H_o}{H_u}=H_o/ t$ and $v:=\frac{a_o^3}{a_o^3+a_u^3}$.
\newline\newline
The average density and its corresponding density parameter are given by:
\begin{equation}
\left\langle \rho\right\rangle_{2region}:=\frac{\rho_oa_o^3+\rho_ua_u^3}{a_o^3+a_u^3} =\frac{\rho_oa_o^3}{a_o^3+a_u^3} =\frac{3}{4\pi G} \frac{\frac{\left(\phi_0-\sin(\phi_0) \right) ^2}{t_0^2\left(1-\cos(\phi) \right) ^3}}{a_o^3+a_u^3}
\end{equation}
\begin{equation}
\Omega_{m}^{2region}:=\frac{8\pi G\left\langle \rho\right\rangle_{2region}}{3H_{2region}^2}
\end{equation}
The final expression for $\left\langle \rho\right\rangle_{2region}$ was obtained from the acceleration equation of the overdense region.
\newline\newline
By utilizing the relations $q_{2region} = \frac{1}{2}\Omega_{m}^{2region}+2\Omega_Q^{2region}$ and $\Omega_m^{2region} + \Omega_{R}^{2region} + \Omega_{Q}^{2region}= 1$, the density parameters of the kinematical backreaction $Q$ and the average curvature can be obtained. Once $\Omega_{R}^{2region}$ is known, $k_{2region}$ can be computed from the definition $k_{2region}(t):=\frac{1}{6}\left\langle ^{(3)}R\right\rangle_{2region}a_{2region}^2 = -\Omega_{R}^{2region}H_{2region}^2a_{2region}^2$.


\begin{thebibliography}{99}
\bibitem{lsst} LSST Science Collaboration, LSST Science Book, Version 2.0, arXiv:0912.0201.

\bibitem{euclid} R. Laureijs et al., Euclid Definition Study Report, arXiv:1110.3193v1 [astro-ph.CO]

\bibitem{realtime} Claudia Quercellini, Luca Amendola, Amedeo Balbi, Paolo Cabella, Miguel Quartin, Real-time Cosmology, Physics Reports, Volume 521, Issue 3, p. 95-134 (2012), arXiv:1011.2646v2 [astro-ph.CO]

\bibitem{parallax1} W.H. McCrea, Observable relations in relativistic cosmology, Zeitschrift fur Astrophysik 9 (1935) 290

\bibitem{parallax} Claudia Quercellini, Miguel Quartin, Luca Amendola, Cosmic Parallax: possibility of detecting anisotropic expansion of the universe by very accurate astrometry measurements, Phys.Rev.Lett.102:151302,2009,  arXiv:0809.3675v3 [astro-ph]

\bibitem{Sandage} Allan Sandage, The Change of Redshift and Apparent Luminosity of Galaxies due to the Deceleration of Selected Expanding Universes, Astrophysical Journal, vol. 136, p.319 (1096)

\bibitem{McVittie} G. C. McVittie, Appendix to The Change of Redshift and Apparent Luminosity of Galaxies due to the Deceleration of Selected Expanding Universes, Astrophysical Journal, vol. 136, p.334 (1962)

\bibitem{realtime_cmb} Raul Jimenez et al., Measuring the Homogeneity of the Universe, arXiv:1902.11298v1 [astro-ph.CO] 

\bibitem{liske} J. Liske et al., Cosmic dynamics in the era of Extremely Large Telescopes, Mon. Not. R. Astron. Soc. 000, 1-27 (2007), arXiv:0802.1532

\bibitem{5years} Stephen Eikenberry et al., Astro2020 White Paper: A Direct Measure of Cosmic Acceleration, arXiv:1904.00217v1 [astro-ph.CO]

\bibitem{codex} L. Pasquini et al., CODEX: Measuring the Expansion of the Universe (and beyond), The Messenger 122, 10 (2005)

\bibitem{drift_with_SKA2} F. Aharonian et al., Pathway to the Square Kilometre Array - The German White Paper -, arXiv:1301.4124v1 [astro-ph.IM]

\bibitem{drift_with_SKA} H.-R. Klockner, D. Obreschkow, C. Martins, A. Raccanelli, D. Champion, A. Roy, A. Lobanov, J. Wagner, R. Keller, Real time cosmology - A direct measure of the expansion rate of the Universe, arXiv:1501.03822v1 [astro-ph.CO]

\bibitem{ska_like} Hao-Ran Yu, Tong-Jie Zhang, Ue-Li Pen, Method for Direct Measurement of Cosmic Acceleration by 21-cm Absorption Systems, Phys. Rev. Lett. 113, 041303, 2014, arXiv:1311.2363v3 [astro-ph.CO] 

\bibitem{fast} Kang Jiao, Jian-Chen Zhang, Tong-Jie Zhang, Hao-Ran Yu, Ming Zhu, Di Li, On the track to measure the apparent cosmic acceleration: observation test and forecast on FAST, arXiv:1905.01184v1 [astro-ph.CO] 



\bibitem{forecast1} Pier-Stefano Corasaniti, Dragan Huterer, Alessandro Melchiorri, Exploring the Dark Energy Redshift Desert with the Sandage-Loeb Test, Phys.Rev.D75:062001,2007, arXiv:astro-ph/0701433v1
\bibitem{forecast2} Matteo Martinelli, Stefania Pandolfi, C. J. A. P. Martins, P. E. Vielzeuf, Probing dark energy with redshift-drift, Phys. Rev. D 86, 123001 (2012), arXiv:1210.7166v3 [astro-ph.CO]
\bibitem{forecast3} Jia-Jia Geng, Jing-Fei Zhang, Xin Zhang, Quantifying the impact of future Sandage-Loeb test data on dark energy constraints, JCAP 07 (2014) 006, arXiv:1404.5407v3 [astro-ph.CO]
\bibitem{forecast4} Jia-Jia Geng, Jing-Fei Zhang, Xin Zhang, Parameter estimation with Sandage-Loeb test, JCAP 12 (2014) 018, arXiv:1407.7123v2 [astro-ph.CO]
\bibitem{forecast5} Shuo Yuan, Siqi Liu, Tong-Jie Zhang, Breaking through the high redshift bottleneck of Observational Hubble parameter Data: The Sandage-Loeb signal Scheme, JCAP02(2015)025, arXiv:1311.1583v4 [astro-ph.CO]
\bibitem{forecast6} Alex G. Kim, Eric V. Linder, Jerry Edelstein, David Erskine, Giving Cosmic Redshift Drift a Whirl, Astroparticle Physics 62, 195 (2015), arXiv:1402.6614v2 [astro-ph.CO]
\bibitem{forecast7} Ming-Jian Zhang, Jing-Zhao Qi, Wen-Biao Liu, Cosmic Expansion History from the Distance Indicator and Redshift Drift, Int J Theor Phys (2015) 54:2456-2466
\bibitem{forecast8} Rui-Yun Guo, Xin Zhang, Constraining dark energy with Hubble parameter measurements: an analysis including future redshift-drift observations, Eur. Phys. J. C 76 (2016) 163, arXiv:1512.07703v4 [astro-ph.CO]
\bibitem{forecast9} Ruth Lazkoz, Iker Leanizbarrutia, Vincenzo Salzano, Forecast and analysis of the cosmological redshift drift, Eur. Phys. J. C (2018) 78: 11, arXiv:1712.07555v1 [astro-ph.CO]

\bibitem{timescape_observables} David L. Wiltshire, Average observational quantities in the timescape cosmology, Phys.Rev.D80:123512,2009, arXiv:0909.0749v2 [astro-ph.CO]

\bibitem{varying_c} Adam Balcerzak, Mariusz P. Dabrowski, Redshift drift in varying speed of light cosmology, Phys. Lett. B728, 15-18 (2014), arXiv:1310.7231v2 [astro-ph.CO]


\bibitem{Copernican} Jean-Philippe Uzan, Chris Clarkson, George F.R. Ellis, Time drift of cosmological redshifts as a test of the Copernican principle, Phys.Rev.Lett.100:191303,2008, arXiv:0801.0068v2 [astro-ph]
\bibitem{void} Chul-Moon Yoo, Tomohiro Kai, Ken-ichi Nakao, Redshift Drift in LTB Void Universes, Phys.Rev.D83:043527,2011, arXiv:1010.0091v1 [astro-ph.CO]
\bibitem{void_mimic} Peter Dunsby, Naureen Goheer, Bob Osano, Jean-Philippe Uzan, How close can an Inhomogeneous Universe mimic the Concordance Model?, JCAP06(2010)017, arXiv:1002.2397v1 [astro-ph.CO]
\bibitem{axial} Priti Mishra, Marie-Noelle Celerier, Tejinder P. Singh, Redshift drift in axially symmetric quasi-spherical Szekeres models, Phys. Rev. D 86, 083520 (2012), arXiv:1206.6026v2 [astro-ph.CO]

\bibitem{Loeb} Abraham Loeb, Direct Measurement of Cosmological Parameters from the Cosmic Deceleration of Extragalactic Objects, Astrophys.J.499:L111-L114,1998, arXiv:astro-ph/9802122v1


\bibitem{variance} Jean-Philippe Uzan, Francis Bernardeau, Yannick Mellier, Time drift of cosmological redshifts and its variance, Phys.Rev.D77:021301,2008, arXiv:0711.1950v2 [astro-ph] 
\bibitem{peculiar_acc1} Luca Amendola, Claudia Quercellini, Amedeo Balbi, Peculiar acceleration, Phys.Lett.B660:81-86,2008, arXiv:0708.1132v1 [astro-ph] 
\bibitem{peculiar_acc2} Claudia Quercellini, Luca Amendola, Amedeo Balbi, Mapping the galactic gravitational potential with peculiar acceleration,  Mon.Not.Roy.Astron.Soc.391:1308-1314,2008 , arXiv:0807.3237v2 [astro-ph]


\bibitem{bc_review1} Thomas Buchert, Syksy Rasanen, Backreaction in late-time cosmology, Annual Review of Nuclear and Particle Science 62 (2012) 57-79, arXiv:1112.5335v2 [astro-ph.CO]
\bibitem{bc_review2} Chris Clarkson et al., Does the growth of structure affect our dynamical models of the universe? The averaging, backreaction and fitting problems in cosmology, Rept.Prog.Phys. 74 (2011) 112901, arXiv:1109.2314v1 [astro-ph.CO]
\bibitem{bc_review3} George F R Ellis, Inhomogeneity effects in Cosmology, Class. Quantum Grav. 28 164001, arXiv:1103.2335v1 [astro-ph.CO]
\bibitem{bc_review4} G.F.R. Ellis, T. Buchert, The universe seen at different scales, Phys.Lett. A347 (2005) 38-46, arXiv:gr-qc/0506106v2

\bibitem{is_bc_important} T. Buchert et al., Is there proof that backreaction of inhomogeneities is irrelevant in cosmology?, Class. Quantum Grav. 32 (2015) 215021, arXiv:1505.07800v2 [gr-qc]


\bibitem{copernican} Chris Clarkson, Bruce A. Bassett, Teresa Hui-Ching Lu, A general test of the Copernican Principle, Phys.Rev.Lett.101:011301,2008, arXiv:0712.3457v2 [astro-ph] 
\bibitem{sum_rule} Syksy Rasanen, Krzysztof Bolejko, Alexis Finoguenov, New test of the FLRW metric using the distance sum rule, Phys. Rev. Lett. 115, 101301 (2015), arXiv:1412.4976v3 [astro-ph.CO]
\bibitem{parallax_test} Syksy Rasanen, A covariant treatment of cosmic parallax, JCAP03(2014)035, arXiv:1312.5738v3 [astro-ph.CO] 




\bibitem{dig_zdrift} S. M. Koksbang, S. Hannestad, Redshift drift in an inhomogeneous universe: averaging and the backreaction conjecture, JCAP01(2016)009, arXiv:1512.05624v2 [astro-ph.CO] 
\bibitem{av_obs2} Syksy Rasanen, Light propagation in statistically homogeneous and isotropic universes with general matter content, JCAP 1003:018,2010, arXiv:0912.3370v2 [astro-ph.CO]

\bibitem{melia} Fulvio Melia, Definitive Test of the $R_h=ct$ Universe Using Redshift Drift, MNRAS Letters 463, (1) L61 (2016), arXiv:1608.00047v1 [astro-ph.CO]


\bibitem{fluid1} Thomas Buchert, On average properties of inhomogeneous fluids in general relativity I: dust cosmologies, Gen.Rel.Grav. 32 (2000) 105-125, arXiv:gr-qc/9906015v2 
\bibitem{bc_fluidII} Thomas Buchert, On average properties of inhomogeneous fluids in general relativity II: perfect fluid cosmologies,  Gen.Rel.Grav.33:1381-1405,2001 , arXiv:gr-qc/0102049v2

\bibitem{light_cone_De1} Ido Ben-Dayan, Maurizio Gasperini, Giovanni Marozzi, Fabien Nugier, Gabriele Veneziano, Do stochastic inhomogeneities affect dark-energy precision measurements?, Phys. Rev. Lett. 110, 021301 (2013), arXiv:1207.1286v2 [astro-ph.CO]
\bibitem{light_cone_De2} I. Ben-Dayan, M. Gasperini, G. Marozzi, F. Nugier, G. Veneziano, Average and dispersion of the luminosity-redshift relation in the concordance model, JCAP06(2013)002, arXiv:1302.0740v3 [astro-ph.CO]


\bibitem{pert1} Daniel Baumann, Alberto Nicolis, Leonardo Senatore, Matias Zaldarriaga, Cosmological Non-Linearities as an Effective Fluid, JCAP07(2012)051arXiv:1004.2488v1 [astro-ph.CO]
\bibitem{pert2} Stephen R. Green, Robert M. Wald, A new framework for analyzing the effects of small scale inhomogeneities in cosmology, Phys.Rev.D83:084020,2011, arXiv:1011.4920v2 [gr-qc]
 
\bibitem{near-FLRW} Syksy Rasanen, Light propagation and the average expansion rate in near-FRW universes, Phys. Rev. D 85, 083528 (2012), arXiv:1107.1176v2 [astro-ph.CO] 
\bibitem{light_cone_bc} Ido Ben-Dayan, Maurizio Gasperini, Giovanni Marozzi, Fabien Nugier, Gabriele Veneziano, Backreaction on the luminosity-redshift relation from gauge invariant light-cone averaging, JCAP04 (2012) 036, arXiv:1202.1247v3 [astro-ph.CO] 
 
\bibitem{multiscale} Alexander Wiegand, Thomas Buchert, Multiscale cosmology and structure-emerging Dark Energy: A plausibility analysis,   Phys.Rev.D82:023523,2010 , arXiv:1002.3912v2 [astro-ph.CO]

\bibitem{Nbody1} Boudewijn F. Roukema, Replacing dark energy by silent virialisation, A\&A 610, A51 (2018), arXiv:1706.06179v2 [astro-ph.CO]
\bibitem{Nbody2} Gábor Racz, Laszlo Dobos, Robert Beck, Istvan Szapudi, Istvan Csabai, Concordance cosmology without dark energy, Mon Not R Astron Soc Lett 2017 slx026, arXiv:1607.08797v2 [astro-ph.CO]
\bibitem{Nbody3} Boudewijn F. Roukema, Jan J. Ostrowski, Thomas Buchert, Virialisation-induced curvature as a physical explanation for dark energy, Journal of Cosmology and Astroparticle Physics 10 (2013) 043, arXiv:1303.4444v3 [astro-ph.CO]

\bibitem{relsim1} Julian Adamek, Chris Clarkson, Ruth Durrer, Martin Kunz, Does small scale structure significantly affect cosmological dynamics?, Phys. Rev. Lett. 114, 051302 (2015), arXiv:1408.2741v2 [astro-ph.CO]
\bibitem{relsim2} Eloisa Bentivegna, Marco Bruni, Effects of nonlinear inhomogeneity on the cosmic expansion with numerical relativity, Phys. Rev. Lett. 116, 251302 (2016), arXiv:1511.05124v3 [gr-qc]
\bibitem{relsim3} John T. Giblin Jr, James B. Mertens, Glenn D. Starkman, Departures from the FLRW Cosmological Model in an Inhomogeneous Universe: A Numerical Examination, Phys. Rev. Lett. 116, 251301 (2016), arXiv:1511.01105v3 [gr-qc]
\bibitem{relsim4} Hayley J. Macpherson, Daniel J. Price, Paul D. Lasky, Einstein's Universe: Cosmological structure formation in numerical relativity, Phys. Rev. D 99, 063522 (2019), arXiv:1807.01711v2 [astro-ph.CO]

\bibitem{fitting1} Julian Adamek, Chris Clarkson, David Daverio, Ruth Durrer, Martin Kunz, Safely smoothing spacetime: backreaction in relativistic cosmological simulations, Class. Quantum Grav. 36, 014001 (2019), arXiv:1706.09309v2 [astro-ph.CO]

\bibitem{peak} Syksy Rasanen, Evaluating backreaction with the peak model of structure formation, JCAP 0804:026,2008, arXiv:0801.2692v3 [astro-ph]

\bibitem{ellipsoid} Francesco Montanari, Syksy Rasanen, Evaluating backreaction with the ellipsoidal collapse model, JCAP12(2017)008, arXiv:1710.02451v2 [astro-ph.CO]




\bibitem{av_obs1} Syksy Rasanen, Light propagation in statistically homogeneous and isotropic dust universes, JCAP 0902:011,2009, arXiv:0812.2872v2 [astro-ph]

\bibitem{template1} Julien Larena et al., Testing backreaction effects with observations, Phys.Rev.D79:083011,2009, arXiv:0808.1161v2 [astro-ph]
\bibitem{template2} Eran Rosenthal, Eanna E. Flanagan, Cosmological backreaction and spatially averaged spatial curvature, arXiv:0809.2107v1 [gr-qc]
\bibitem{template3} Aseem Paranjape, T. P. Singh, Explicit Cosmological Coarse Graining via Spatial Averaging, Gen.Rel.Grav.40:139-157,2008, arXiv:astro-ph/0609481v4



\bibitem{boehm} Celine Boehm, Syksy Rasanen, Violation of the FRW consistency condition as a signature of backreaction, JCAP09(2013)003, arXiv:1305.7139v3 [astro-ph.CO]



\bibitem{Syksy_2region} Syksy Rasanen, Accelerated expansion from structure formation, JCAP0611:003,2006, arXiv:astro-ph/0607626v3
\bibitem{Syksy_2region2} Syksy Rasanen, Cosmological acceleration from structure formation, Int.J.Mod.Phys.D15:2141-2146,2006, arXiv:astro-ph/0605632v1
\bibitem{Syksy_2region3} Syksy Rasanen, Backreaction: directions of progress, Class. Quantum Grav. 28 (2011) 164008, arXiv:1102.0408v2 [astro-ph.CO]


\bibitem{Bull_tworegion} Philip Bull, Timothy Clifton, Local and non-local measures of acceleration in cosmology, Phys. Rev. D 85, 103512 (2012), arXiv:1203.4479v2[astro-ph.CO]
\bibitem{multiscale2} Thomas Buchert, Mauro Carfora, On the curvature of the present-day Universe, Class.Quant.Grav.25:195001,2008, arXiv:0803.1401v2 [gr-qc]

\bibitem{use2region} Celine Boehm, Syksy Rasanen, Violation of the FRW consistency condition as a signature of backreaction, JCAP09(2013)003, arXiv:1305.7139v3 [astro-ph.CO]

\bibitem{arbitrary_spacetime} Stella Seitz, Peter Schneider, Jurgen Ehlers, Light propagation in arbitrary spacetimes and the gravitational lens approximation, Class.Quant.Grav.11:2345-2374,1994 , astro-ph/9403056

\bibitem{Tardis} Mikko Lavinto, Syksy Rasanen, Sebastian J. Szybka, Average expansion rate and light propagation in a cosmological Tardis spacetime, JCAP12(2013)051, arXiv:1308.6731v2 [astro-ph.CO]

\bibitem{marie-noelle_sn} Marie-Noelle Celerier, Do we really see a cosmological constant in the supernovae data?, Astron.Astrophys.353:63-71,2000, arXiv:astro-ph/9907206v4

\bibitem{rlp} Andrzej Krasinski, Krzysztof Bolejko, Redshift propagation equations in the $\beta' \neq0$  Szekeres models, Phys.Rev.D83:083503,2011, arXiv:1007.2083v2 [gr-qc]



\bibitem{fitting3} Stephen R. Green, Robert M. Wald, How well is our universe described by an FLRW model?, Class. Quantum Grav. 31 (2014) 234003, arXiv:1407.8084v2 [gr-qc] 
\bibitem{lightcone_dz} M. Gasperini, G. Marozzi, F. Nugier, G. Veneziano, Light-cone averaging in cosmology: formalism and applications, JCAP 1107:008,2011, arXiv:1104.1167v3 [astro-ph.CO]

\bibitem{fitting2} Valerio Marra, Edward W. Kolb, Sabino Matarrese, Light-cone averages in a swiss-cheese universe, Phys.Rev.D77:023003,2008, arXiv:0710.5505v2 [astro-ph]


\bibitem{linder} Eric V. Linder, Averaging Inhomogeneous Universes: Volume, Angle, Line of Sight, arXiv:astro-ph/9801122v2

\bibitem{dig_selv_tardis} S. M. Koksbang, Towards statistically homogeneous and isotropic perfect fluid universes with cosmic backreaction, Sofie Marie Koksbang 2019 Class. Quantum Grav. in press https://doi.org/10.1088/1361-6382/ab376c, arXiv:1907.08681v1 [gr-qc] 

\bibitem{virialisation} Boudewijn F. Roukema, Jan J. Ostrowski, Thomas Buchert: Virialisation-induced curvature as a physical explanation for dark energy, Journal of Cosmology and Astroparticle Physics 10 (2013) 043, arXiv:1303.4444v3 [astro-ph.CO]


\bibitem{teppo1} Maria Mattsson, Teppo Mattsson: On the role of shear in cosmological averaging,  10.1088/1475-7516/2010/10/021 , arXiv:1007.2939v1 [astro-ph.CO]
\bibitem{teppo2} Maria Mattsson, Teppo Mattsson: On the role of shear in cosmological averaging II: large voids, non-empty voids and a network of different voids, 10.1088/1475-7516/2011/05/003, arXiv:1012.4008v1 [astro-ph.CO]

\bibitem{dig_CMB} S. M. Koksbang, Light propagation in Swiss cheese models of random close-packed Szekeres structures: Effects of anisotropy and comparisons with perturbative results, Phys. Rev. D 95, 063532 (2017), arXiv:1703.03572v2 [astro-ph.CO]





\bibitem{obs_coord} Roy Maartens, Neil Humphreys, David Matravers, Bill Stoeger, Inhomogeneous universes in observational coordinates, Class.Quant.Grav. 13 (1996) 253-264; Erratum-ibid. 13 (1996) 1689, arXiv:gr-qc/9511045v1

\bibitem{lc_not_obs} G. Fanizza, M. Gasperini, G. Marozzi, G. Veneziano: An exact Jacobi map in the geodesic light-cone gauge, JCAP 11 (2013) 019, arXiv:1308.4935v2 [astro-ph.CO]

\bibitem{Ryden} Barbara Ryden, Introduction to cosmology, OPearson Education Int., publishing as Addison Wesley, ISBN 0-8053-8912-1, 2003









\end{thebibliography}
\end{document}